\documentclass[a4paper,12pt]{article}

\usepackage{amsmath}
\usepackage{latexsym}
\usepackage{cite}
\usepackage{graphicx}
\usepackage[small,bf,hang]{caption}

\newcommand{\cg}{C(G)}

\newcommand{\fdual}{\tilde{F}}

\newcommand{\first}{\ensuremath{\mbox{1}^{\mbox{\tiny st}}}}

\newcommand{\medsp}{\\[0.7ex]}

\newcommand{\second}{\ensuremath{\mbox{2}^{\mbox{\tiny nd}}}}

\newcommand{\Lindex}[1]{\ensuremath{\smallindex{\mathcal{L}}{#1}}}
\newcommand{\Ltext}[1]{\ensuremath{\itindex{\mathcal{L}}{#1}}}

\newcommand{\bra}[1]{\langle #1 |}

\newcommand{\diff}[1]{\mbox{d}#1}

\newcommand{\elim}[1]{\itindex{#1}{elim}}
\newcommand{\eff}[1]{\itindex{#1}{eff}}
\newcommand{\half}[1]{\ensuremath{\frac{#1}{2}}}
\newcommand{\intd}[1]{\int \!\! #1 \;}
\newcommand{\inv}[1]{\ensuremath{\frac{1}{#1}}}
\newcommand{\ket}[1]{| #1 \rangle}
\newcommand{\lomega}[1][]{\bra{\Omega #1}}
\newcommand{\metr}[1][]{g_{\varphi \bar{\varphi} #1}}

\newcommand{\romega}[1][]{\ket{\Omega #1}}

\newcommand{\itindex}[2]{\ensuremath{#1_{\mbox{\scriptsize{\itshape #2}}}}} 
\newcommand{\smallindex}[2]{\ensuremath{#1_{\scriptscriptstyle{#2}}}}

\newcommand{\varfrac}[2][]{\frac{\delta #1}{\delta #2}}

\DeclareMathOperator{\tr}{Tr}
\DeclareMathOperator{\hc}{h.c.}

\begin{document}
%

\renewcommand{\thefootnote}{\fnsymbol{footnote}}
\thispagestyle{empty}
\begin{titlepage}

\vspace*{-1.8cm}
\hfill \parbox{3.5cm}{hep-th/0312034\\TUW-03-36}
\vspace*{2cm}

\begin{center}
  {\Large {\bf \hspace*{-0.2cm} No Supersymmetry without
      Supergravity\protect\vspace*{0.3cm}\\ {\large Induced Supersymmetry Representations on \protect\vspace*{0.3cm}\\
      Composite Effective Superfields}}}
      \vspace*{1cm} \\

{\bf L. Bergamin\footnote{email: bergamin@tph.tuwien.ac.at, phone: +43  1 58801 13622, fax: +43 1 58801 13699}}\\
      Institute for Theoretical Physics \\
    Technical University of Vienna \\
    Wiedner Hauptstr.\ 8-10 \\
    A-1040 Vienna, Austria\\[3ex]
{\bf P. Minkowski\footnote{email: mink@itp.unibe.ch, phone: +41 31 631 8624,
    fax: +41 31 631 3821}}\\
    Institute for Theoretical Physics \\
    University of Bern \\
    Sidlerstrasse 5\\
    CH - 3012 Bern, Switzerland
   \vspace*{0.8cm} \\  



\begin{abstract}
\noindent
Induced supersymmetry representations on composite operators are studied. In
superspace the ensuing transformation rules (trivially) lead to an effective
superfield. On the other hand, an
induced representation must exist for non-linear (``on-shell'') supersymmetry
as well. As this choice of the representation is physically irrelevant,
any formulation of an effective action
starting from the superspace representation must equally well
be possible in a non-linear representation. We show that this leads to very
relevant constraints on the formulation of effective actions in terms of
composite operators. These ideas are applied to the simplest case of such a theory, $N=1$
SYM. It is shown that soft supersymmetry breaking within that theory forces
one to include besides the Lagrangian multiplet $S$ all currents
of the super-conformal structure, embedded in a supergravity background, as relevant fields. 

\vspace{3mm}
\noindent
{\footnotesize {\it PACS:}  11.15.Kc; 11.30.Pb; 11.30.Qc \newline
{\it keywords:} SYM theory; non-linear representations; low-energy effective action}
\end{abstract}
\end{center}

\end{titlepage}

\renewcommand{\thefootnote}{\arabic{footnote}}
\setcounter{footnote}{0}
\numberwithin{equation}{section}
%

\section{Introduction}
$N=1$ SYM theories attracted much attention during the last year. On the one
hand this is due to substantial progress in the understanding of
supersymmetric gauge theories including $N=1$ SYM as a sub-sector, based on the
study of string theory/matrix models as well as field-theoretic methods 
\cite{Dijkgraaf:2002fc,Dijkgraaf:2002vw,Dijkgraaf:2002dh,Dijkgraaf:2002xd,Ferrari:2002jp,Ferrari:2002kq,Ferrari:2003yr,Cachazo:2002ry,Cachazo:2002zk,Cachazo:2003yc}. On the other
hand first results from lattice simulations are available \cite{Peetz:2002sr,Feo:2002yi,Feo:2003jq}, which should allow
tests of analytic results in the near future. However, there
could arise serious difficulties in the comparison of the two approaches to
SYM theories: Mostly analytic calculations are restricted to the study of the
superpotential, as its holomorphic dependence is one of the most important constraints
within supersymmetric theories. Also the set of physical fields considered
therein \cite{veneziano82,Shore:1983kh} might not be general enough: In most
applications there appear only the gluino bilinear operator  with its
fermionic partner generating the
low energy spectrum.

In two recent papers \cite{bergamin02:1,Bergamin:2003ub} the authors have
shown that the restriction to determine the superpotential and the difficulty
of the missing
glue-balls  are actually related. Two operators generating
glue-balls, $\tr F_{\mu \nu} F^{\mu \nu}$ and $\tr F_{\mu \nu} \fdual^{\mu \nu}$,
are part of the effective superfield $S \propto \tr W^{\alpha} W_{\alpha}$, but
they appear as \emph{highest} component, i.e.\ at a place where one usually
expects an auxiliary field. As long as the explicit calculations are
restricted to the superpotential $W(S)$, this ``auxiliary'' field is not
integrated out explicitely, as this step would require detailed knowledge
about the non-holomorphic part of the effective
Lagrangian. Nevertheless, it is assumed \emph{implicitely} in most cases, that
the highest component of $S$ is auxiliary: If the first derivatives $W_{,
  i}$ of the
superpotential shall define the minimum and the second derivative $W_{,
 ij}$ the
spectrum, then all highest components of the involved chiral fields \emph{must
be standard auxiliary fields}. Thus the glue-ball pair-operators drop out of the
action\footnote{Contrary to this assumption $S$ is often called ``glue-ball superfield''.}.

We briefly want to recall why this step cannot be consistent from fundamental
physical considerations. In any supersymmetric theory formulated in superspace
there appear auxiliary fields. E.g.\ in the Wess-Zumino model the highest
component of the chiral field $\itindex{\Phi}{WZ} = \varphi + \theta \psi + \theta^2 \itindex{F}{WZ}$
obeys
\begin{equation}
  \label{eq:intro1}
  \lomega T \itindex{F}{WZ}(x) \itindex{F}{WZ}(y) \romega \propto \delta(x-y)\ .
\end{equation}
This relation holds in the full quantum theory as a consequence of
supersymmetry Ward-identities. Thus the action of the Wess-Zumino model is ultra-local in the
classical as well as in the quantum theory and, as a consequence,
$\itindex{F}{WZ}$ may be eliminated without changing the physics of the
model\footnote{Here we consider the renormalizable linear model, only. In that
case the integration over the auxiliary field in the path integral does not
lead to a non-trivial functional determinant. Of course, this is not true in a
non-linear model and thus algebraic equations of motion are not sufficient to
ensure at the
quantum level the equivalence of a theory with and without auxiliary fields,
resp. However, the present discussions about the interpretation of auxiliary
fields does not concern this complication.}.

In its confined region, $N=1$ SYM theory is described by a chiral field $S \propto \tr W^\alpha W_\alpha = \varphi +
\theta \psi + \theta^2 \itindex{F}{SYM}$ as
well, the Lagrangian multiplet. This is
not a fundamental field, but a superfield of composite operators. Thus the
two-point function of $\itindex{F}{SYM}$
\begin{multline}
  \label{eq:intro2}
  \lomega T \itindex{F}{SYM}(x) \itindex{F}{SYM}(y) \romega\medsp
  \propto \lomega T
  (\tr F_{\mu \nu} F^{\mu \nu} - i F_{\mu \nu} \fdual^{\mu \nu} )(x) (\tr
  F_{\mu \nu} F^{\mu \nu} - i F_{\mu \nu} \fdual^{\mu \nu} )(y) \romega + \ldots \ .
\end{multline}
is fundamentally different from
\eqref{eq:intro1}: \eqref{eq:intro2} is the two-point function of dynamical
variables, the glue-ball operators. It cannot be ultra-local as \eqref{eq:intro1} and
in fact supersymmetry does not at all dictate this. 

Therefore the relevant question is not, which operators represent the
glue-balls in SYM, but how the glue-ball operators in $\itindex{F}{SYM}$ can
appear in an effective action \emph{dynamically}. Based on an extension of the Veneziano-Yankielowicz Lagrangian first proposed
in ref.\ \cite{Shore:1983kh}, the present
authors showed in refs.\ \cite{bergamin02:1,Bergamin:2003ub,Bergamin:2003xc} how to achieve this. The resulting action has four bosonic degrees of
freedom from $\varphi$ and $F$, but only two fermionic ones. The purpose of
the present paper is to illuminate this mismach by the study of the induced
supersymmetry transformations of $S$ in both, the superspace representation as
well as the Wess-Zumino gauge. It is found, that a resolution requires a much larger context, including the whole
superconformal structure embedded in a supergravity background.
 
\section{An Example: The Wess-Zumino Model}
\label{sec:wzexample}
To clarify the important difference in the symmetry content of single and
composite effective fields resp., the basic ideas are worked out for the
Wess-Zumino model in a first step. The absence of a gauge-symmetry drastically
simplifies the technical aspects, but at the same time all relevant issues are
still reproduced. Nevertheless, this section should be seen as an exercise
preparing the more complicated case of SYM. We do not claim that all steps
performed here have the same meaning within the Wess-Zumino
model.
\subsection{Simple Effective Fields}
We consider the Wess-Zumino model with a single chiral field $\Phi = \varphi
+ \theta \psi + \theta^2 F$ and the Lagrangian
\begin{equation}
  \label{eq:wzmodel}
  \Ltext{Wess-Zumino} = \intd{\diff{^4 \theta}} \bar{\Phi} \Phi - \bigl(
  \intd{\diff{^2 \theta}} \half{m} \Phi^2 + \frac{\lambda}{3} \Phi^3 +
  \hc\bigr)\ .
\end{equation}
The effective action may be written in terms of the simple low-energy fields
$\eff{\Phi} = \lomega[(J)] \Phi \romega[(J)]$, which is obtained from the
source extension of the action
\begin{equation}
  \label{eq:wzsimplesource}
  \Lindex{J} = \Ltext{Wess-Zumino} + \bigl( \intd{\diff{^2 \theta}} J \Phi +
  \hc \bigr)
\end{equation}
and the subsequent Legendre transformation\footnote{All low energy actions in this work are defined as associated limits
    of effective actions, valid over the entire range of momenta and 
    externally prescribed field. However, the main
  conclusions do not rely on this definition, but only on the
  fact that a supersymmetry covariant set of effective operators may be
  defined as those, which couple to a source term (local coupling). This
  applies to any other low energy approximation (as e.g. the Wilsonian action) as
  well, though no Legendre transformation is performed in these cases.}
\begin{equation}
  \label{eq:wzGamma1}
  \Gamma[\eff{\Phi},\eff{\bar{\Phi}}] = \intd{\diff{^4x}}
  \bigl(\intd{\diff{^2 \theta}} J \eff{\Phi} + \hc \bigr) - W[J,\bar{J}]\ .
\end{equation}
By the definition \eqref{eq:wzGamma1} the effective action is a integral over
superspace as well and thus takes the form\footnote{Of course, holomorphicity
  arguments in that particular case reduce the effective superpotential to its
classical form. As this simplification could hide the relevant observations,
we keep a general effective superpotential in the following.}
\begin{equation}
  \label{eq:wzs1}
  \Gamma[\eff{\Phi},\eff{\bar{\Phi}}] = \intd{\diff{^4x}} \biggl(\intd{\diff{^4 \theta}}
  K(\eff{\Phi}, \eff{\bar{\Phi}})
  - \bigl( \intd{\diff{^2 \theta}} W(\eff{\Phi}) + \hc \bigr)
  \biggr)\ .
\end{equation}
The classical action \eqref{eq:wzmodel} is invariant under the supersymmetry
transformations
\begin{align}
  \label{eq:wzs2}
  \delta \varphi &= \epsilon \psi\ , & \delta \psi_\alpha &= \epsilon_\alpha F -
  i (\sigma^\mu \bar{\epsilon})_\alpha \partial_\mu \varphi\ , & \delta F &= -i
  \bar{\epsilon} \bar{\sigma}^\mu \partial_\mu \psi
\end{align}
and they are inherited by the effective fields
\begin{align}
    \label{eq:wzs3}
  \delta \eff{\varphi} &= \epsilon \eff{\psi}\ , & \delta {\eff{\psi}}_\alpha &= \epsilon_\alpha \eff{F} -
  i (\sigma^\mu \bar{\epsilon})_\alpha \partial_\mu \eff{\varphi}\ , & \delta \eff{F} &= -i
  \bar{\epsilon} \bar{\sigma}^\mu \partial_\mu \eff{\psi}\ .
\end{align}
Supersymmetry covariance of the source extension is manifest as the action
\eqref{eq:wzsimplesource} is invariant under that symmetry if the components
of $J = j + \theta \eta + \theta^2 f$ transform as a chiral field as well.

On the classical level the masses of $\varphi$ and $\psi$ are degenerate. It
is important to realize that this does not follow from the superpotential
alone, but the full potential
\begin{equation}
  \label{eq:wzs4}
  V = - \bar{F} F + \bigl( m(F \varphi - \half{1} \psi \psi) + \lambda (F
  \varphi^2 - \varphi \psi \psi) + \hc \bigl) - (j F + f \varphi - \eta \psi +
  \hc)
\end{equation}
is important. Only after elimination of the auxiliary field
\begin{equation}
  \label{eq:wzs5}
  \elim{F} = \bar{m} \bar{\varphi} + \bar{\lambda} \bar{\varphi}^2 - \bar{j}
\end{equation}
the mass of the scalar field becomes manifest. The superpotential alone as a
holomorphic function must have unstable directions in the complex fields
$\varphi$ and $F$. Therefore the masses of the theory are not defined without
detailed knowledge of the K\"{a}hler potential.

After the elimination of $F$ the symmetry transformations \eqref{eq:wzs2} become (for $J = 0$)
\begin{align}
  \label{eq:wzs6}
  \delta \varphi &= \epsilon \psi\ , & \delta \psi_\alpha &= \epsilon_\alpha \elim{F} -
  i (\sigma^\mu \bar{\epsilon})_\alpha \partial_\mu \varphi\ , & \delta
  \elim{F} &= \bar{m} \bar{\epsilon} \bar{\psi} + 2 \bar{\lambda}
  \bar{\epsilon} \bar{\psi} \bar{\varphi} \ .
\end{align}
The transformation of the dependent field $\elim{F}$ is equivalent to $\delta
F$ in \eqref{eq:wzs2} up to equations of motion.

The purpose of the present paper is, to clarify the r\^{o}le of the elimination
of auxiliary fields in the context of effective actions. On the one hand
the classical theory shows already that a similar procedure can be important
for the effective actions as well, namely if $|W_{, \varphi}|^2$  shall determine the
scalar potential. On the other hand, the non-holomorphic part of the effective action
need not be restricted to a simple form as in \eqref{eq:wzmodel}, in
particular there could appear derivative terms on the ``effective auxiliary
field'', which promotes the latter to a propagating degree of freedom. To
avoid misunderstandings in the discussion of different types of ``auxiliary''
fields we introduce the following nomenclature \cite{bergamin01}:
\begin{enumerate}
\item The auxiliary fields of the fundamental theory ($F$ in the present case)
  are called \first\ generation or fundamental auxiliary fields.
\item Fields in an effective multiplet that appear at a place where one
  usually expects an auxiliary field are called \second\ generation or effective
  ``auxiliary'' fields. The word ``auxiliary'' is put in quotation marks, as
  these fields need not have the typical behavior of an auxiliary field, but
  can contain propagating modes.
\end{enumerate}

In the first place it is important to realize that \second\ generation auxiliary
fields cannot be eliminated in the sense of \eqref{eq:wzs5}. The number of
effective fields does not depend on the number of fundamental fields, which
changes by the elimination \eqref{eq:wzs5}, but only on the number of
sources. For a chiral field this number is always three, there exists no
concept of ``elimination of a source'', which would maintain the supersymmetry
covariance\footnote{Local coupling constants are a powerful tool to
  investigate supersymmetric theories in $x$-space
  \cite{flume99,Kraus:2001kn,kraus01,kraus01:3}. But as the construction of local
  couplings is closely related to the formulation in superspace, supersymmetry
  is enforced in the enveloping superspace, but not reducible to the
  evaluation of supergraphs.}. The source $j$
seems to disappear from the potential after
elimination of $F$ as
\begin{equation}
  \label{eq:wzs6.1}
  V = \itindex{F}{elim} \itindex{\bar{F}}{elim} - \bigl( (\half{m} + \lambda
  \varphi) \psi \psi + \hc \bigr) - (f \varphi - \eta \psi + \hc)\ .
\end{equation}
But while the action is supersymmetry covariant with all three sources $j$,
$\eta$ and $f$ this is not the case after reducing the sources to $\eta$ and
$f$ only, as \eqref{eq:wzs5} depends on $j$.

Nevertheless, in the context of the present example there \emph{must} exist a
concept of the elimination of the \second\ generation auxiliary field: Indeed,
this field (in this particular example) is just the low energy version of the
fundamental auxiliary field and thus should inherit its fundamental
properties. Indeed, from \eqref{eq:wzs5} follows after the elimination of the
fundamental auxiliary field 
\begin{equation}
  \label{eq:wzs7}
  \eff{F}(x) = \varfrac{j(x)} W[j,\eta,f;\bar{j},\bar{\eta},\bar{f}] = \lomega[(J)]
  \bar{m} \bar{\varphi} + \bar{\lambda} \bar{\varphi}^2 \romega[(J)][(J)]\ ,
\end{equation}
i.e.\ $\eff{F}$ \emph{becomes a function of the
  fields} $\eff{\varphi}$ \emph{and} $\eff{\psi}$. Thus it is necessary and consistent to ``eliminate''
  the effective auxiliary field as well and to postulate
  \begin{equation}
  \label{eq:wzs8}
    \lomega[(J)] \bar{m} \bar{\varphi} + \bar{\lambda} \bar{\varphi}^2 - \bar{j} \romega[(J)]
    \approx \inv{\metr}(\bar{W}_{, \bar{\varphi}} + \half{1} \metr[,
    \bar{\varphi}] \eff{\psi} \eff{\psi})\ ,
  \end{equation}
where $\metr$ is the K\"{a}hler metric of the non-holomorphic part in
\eqref{eq:wzs1}. This ``elimination'' has to be understood such that the
relation
\begin{align}
  \varfrac{j(x)} W[j,\eta,f;\bar{j},\bar{\eta},\bar{f}] &= \eff{F}(x)
\end{align}
may be replaced by a function of variations with respect to the sources
$\eta$, $f$ and their hermitian conjugates. It does not mean that $\eff{F}$
disappears from the theory, as the variation
\begin{align}
  \varfrac{\eff{F}(x)} \Gamma[\eff{\varphi}, \eff{\psi}, \eff{F},
  \eff{\bar{\varphi}}, \eff{\bar{\psi}}, \eff{\bar{F}} ] &= j(x)
\end{align}
must retain its validity.

To conclude, the close relation between the \first\ and the \second\
generation auxiliary fields induces the ``elimination'' of the latter as a
consequence of the elimination of the former using the algebraic equations of motion. This dictates
that the equations of motion for $\eff{F}$ in \eqref{eq:wzs1} are algebraic as
well, i.e.\ $K(\eff{\Phi}, \eff{\bar{\Phi}})$ is a K\"{a}hler manifold. Still,
as outlined above, the elimination of the \first\ generation auxiliary field and
functional restriction of the \second\ generation auxiliary field in terms of the physical
effective fields are two different procedures. This is relevant to understand
the subsequent sections, even if it may appear
hair-splitting.

In the generic case the relation \eqref{eq:wzs8} can be complicated, as
$\lomega \varphi^2 \romega \neq \left(\lomega \varphi \romega \right)^2$ in
general. However, in the present example holomorphicity arguments restrict the
effective superpotential to the classical superpotential, which once again
justifies the above line of arguments.
\subsection{Composite Effective Fields}
The situation changes drastically, once the
source is not coupled to a field monomial, but rather to a composite
operator. As an example we consider the action \eqref{eq:wzmodel} together
with a source extension coupled to $\Phi^2 = \Psi$, only:
\begin{equation}
  \label{eq:wzcompsource}
  \Lindex{J} = \Ltext{Wess-Zumino} + \bigl( \intd{\diff{^2 \theta}} J \Phi^2 +
  \hc \bigr)
\end{equation}
The new effective field $\eff{\Psi} = \eff{\phi} + \theta \eff{\kappa} +
\theta^2 \eff{D}$ follows from variation with respect to the source $J$ in
\eqref{eq:wzcompsource}. The field content of its components reads
\begin{align}
  \label{eq:wzc1}
  \eff{\phi} &= \lomega[(J)] \varphi^2 \romega[(J)]\ , & (\eff{\kappa})_\alpha
  &= \lomega[(J)] 2 \varphi 
  \psi_\alpha \romega[(J)]\ , & \eff{D} &= \lomega[(J)] 2 \varphi F - \psi \psi \romega[(J)]\ .
\end{align}
Of course, the three fields transform as a chiral field as may be tested
easily by applying the transformation rules \eqref{eq:wzs2} to the operators
in \eqref{eq:wzc1}.

As in the previous section the effect of the elimination of the fundamental
auxiliary field on the new effective action\footnote{This effective action is
  defined for illustrative purposes, only. We do not intend that the
  full dynamics of the Wess-Zumino model are correctly reproduced by this
action.}
\begin{equation}
  \label{eq:wzc2}
    \Gamma[\eff{\Psi},\eff{\bar{\Psi}}] = \intd{\diff{^4x}} \biggl(\intd{\diff{^4 \theta}}
  K(\eff{\Psi}, \eff{\bar{\Psi}})
  - \bigl( \intd{\diff{^2 \theta}} W(\eff{\Psi}) + \hc \bigr)
  \biggr)
\end{equation}
must be studied.

From eq.\ \eqref{eq:wzs5} the three operators in
$\Psi$ after elimination of $F$ become ($J = 0$):
\begin{align}
  \label{eq:wzc3}
  \phi &= \varphi^2 & \kappa_\alpha &= 2 \varphi
  \psi_\alpha & \itindex{D}{elim} &=  2 \bar{m} |\varphi|^2 + 2 \bar{\lambda}
  |\varphi|^2 \bar{\varphi} - \psi \psi 
\end{align}
The supersymmetry transformations
\begin{align}
  \label{eq:wzc4}
  \delta \phi &= \epsilon \kappa\ , & \delta \kappa_\alpha &=
  \epsilon_\alpha \itindex{D}{elim} - i (\bar{\sigma}^\mu \bar{\epsilon})_\alpha
  \partial_\mu \phi\ ,
\end{align}
are still functions of the operators\footnote{We denote the set of operators
  $(\phi, \kappa, D)$ and the superfield $\phi + \theta \kappa + \theta^2 D$
  by the same symbol $\Psi$. The meaning should be clear within the context.} in $\itindex{\Psi}{elim}$,
while $\itindex{D}{elim}$ transforms as
\begin{equation}
  \label{eq:wzc5}
  \delta \itindex{D}{elim} = (2 \bar{m} \varphi + 4 \bar{\lambda}
  |\varphi|^2) \bar{\epsilon} \bar{\psi} - 2i \bar{\epsilon} \bar{\sigma}^\mu
  \psi \partial_\mu \varphi\ . 
\end{equation}
Eqs.\ \eqref{eq:wzc4} and \eqref{eq:wzc5} unravel several important differences compared to the case of simple
effective fields of the previous section:

If the effective fields are composite operators of the fundamental ones, then
the elimination of the fundamental auxiliary field does not induce the
elimination of any of the effective fields. Indeed, $\itindex{D}{elim/eff} =
\lomega[(J)] \itindex{D}{elim} \romega[(J)]$ cannot
be written as a function of $\eff{\phi}$ and $\eff{\kappa}$.

If, accidentally, an ansatz for the effective action
$\Gamma[\eff{\Psi},\eff{\bar{\Psi}}]$ had algebraic
equations of motion for $\itindex{D}{elim/eff}$, this ansatz would fail to
capture fundamental properties of the system: Elimination of
$\itindex{D}{elim/eff}$ would lead to an action formulated solely in terms of
$\eff{\phi}$ and $\eff{\kappa}$, but this action cannot be supersymmetric as
supersymmetry never closes on these two fields. Moreover, such an action would
be ultra-local in $\itindex{D}{elim/eff}$, which is easily falsified:
$\itindex{D}{elim/eff}$ represents \emph{propagating} degrees of
freedom.

This leads to the definite conclusion: \emph{If the effective action of a
  supersymmetric theory is formulated in terms of composite operators only, then
  there exist no auxiliary fields among the variables of this action. All
  fields must be interpreted as (independent) propagating modes.}

Yet the present example seems to be inconsistent: Clearly
supersymmetry does not close on the set of operators in $\eff{\Psi}$ once the
fundamental auxiliary field has been eliminated. However, this elimination
should have no influence on the physics. Therefore, in all applications of
composite effective supermultiplets supersymmetry covariance must be restored
due to a complete set of additional relations.

\section{SUSY representation of confined SYM}
The choice of sources for composite operators is important in $N=1$ SYM theory, as this theory exhibits confinement. The
relevant composite superfield $S$ (the Lagrangian or anomaly multiplet) is
best defined through the anomalous current conservation
\begin{align}
  \label{eq:SYM1}
  \bar{D}^{\dot{\alpha}} V_{\alpha \dot{\alpha}} &= \delta_\alpha\ , &
  \delta_\alpha &=- 2 D_\alpha S\ , & \bar{D}_{\dot{\alpha}} S
  &= 0\ , \medsp
  S &= c \tr W^\alpha W_\alpha\ , & c &= -
  \frac{\beta}{24 g^3 \cg}\ . &&
\end{align}
In terms of the physical fields $A_\mu$ (gluon), $\lambda_\alpha$ (gluino) and
the real auxiliary field $D$ the components of $S = \varphi + \theta \psi +
\theta^2 F$ take the form
\begin{align}
\label{eq:SYM1.1}
  \varphi &= 2 c \tr \lambda \lambda \medsp
\label{eq:SYM1.2}
  \psi_\alpha &= \sqrt{2} c \bigl(2 \tr \lambda_\alpha D - \tr (\sigma^{\mu \nu}
  \lambda)_\alpha F_{\mu \nu}  \bigr) \medsp
\label{eq:SYM1.3}
  F &= 4 i c \tr \lambda \sigma^\mu D_\mu \bar{\lambda} - c (\tr F_{\mu \nu}
  F^{\mu \nu} - i \tr F_{\mu \nu} \fdual^{\mu \nu}) + 2 c \tr D^2
\end{align}
The classical Lagrangian is proportional\footnote{In effective calculations
  the definition of $S$ as given in \eqref{eq:SYM1} should be used, as this
  quantity is renormalization group invariant. However, in purely classical
  considerations below the quantity $\tilde{S}$ is used.} to $S$
\begin{align}
  \label{eq:SYM2}
  \Ltext{SYM} &= \intd{\diff{^2 \theta}} \tau_0 \tilde{S} + \hc\ , & S &= -
  \frac{\beta}{3 g^3} \tilde{S}
\end{align}
with the complex coupling constant $\tau_0 = \inv{g^2} + \frac{i \vartheta}{8
  \pi^2}$. An effective action in terms of the field $\eff{S} = \lomega[(J)] S
  \romega[(J)]$ is obtained by the extension of $\tau$ to a chiral superfield $J =
  \tau + \theta \eta - 2 \theta^2 m$ and a subsequent Legendre transformation
  with respect to this source \cite{Shore:1983kh,burgess95,bib:mamink,bergamin01}.

From the supersymmetry transformations\footnote{As all quantities used in this
work are gauge-invariant, there appear no difficulties in the use of
supersymmetry transformations in Wess-Zumino gauge together with the
superspace formulation.}
\begin{align}
  \delta A_\mu &= \inv{\sqrt{2}}(\bar{\epsilon}\bar{\sigma}_\mu\lambda +
  \bar{\lambda} \bar{\sigma}_\mu \epsilon) \medsp
  \delta \lambda_\alpha &= \inv{\sqrt{2}} \epsilon_\alpha D + \inv{2 \sqrt{2}}
  (\sigma^{\mu \nu})_\alpha F_{\mu \nu} \medsp
  \delta D &= - \frac{i}{\sqrt{2}}(\epsilon \sigma^\mu D_\mu \bar{\lambda} +
  \bar{\epsilon} \bar{\sigma^\mu D_\mu \lambda})
\end{align}
it follows that $S$ transforms as a chiral superfield as long as $D$ has not
been eliminated. The same should then apply to the effective field $\eff{S}$
and thus the effective action of SYM takes the form
\begin{equation}
  \label{eq:SYM3}
  \itindex{\Gamma}{SYM} = \intd{\diff{^4 \theta}} K(\eff{S},\eff{\bar{S}}) - \bigl(
  \intd{\diff{^2 \theta}} W(\eff{S}) + \hc \bigr)\ .
\end{equation}
The partially anomalous current conservation \eqref{eq:SYM1} determines the
superpotential \cite{veneziano82,Shore:1983kh,burgess95} $W(\eff{S}) \propto \eff{S} (\log
\eff{S}/\Lambda^3 -1)$, where $\Lambda$ is the --in general complex-- scale of
the theory.

From \eqref{eq:SYM2} with \eqref{eq:SYM1.3} it is easily seen that the
auxiliary field is eliminated by the trivial equation of motion $D = 0$. After
this elimination the new operators
\begin{align}
  \itindex{\varphi}{elim} &= \varphi\ , & \itindex{\psi}{elim} &= - \sqrt{2} c \tr (\sigma^{\mu \nu}
  \lambda) F_{\mu \nu}\ , & \itindex{F}{elim} &= F - 2 c \tr D^2
\end{align}
transform as
\begin{align}
\label{eq:SYM4}
  \delta \itindex{\varphi}{elim} &= \epsilon \itindex{\psi}{elim}\ , \medsp
\label{eq:SYM5}
  \delta (\itindex{\psi}{elim})_\alpha &= \epsilon_\alpha \itindex{F}{elim} -
  i (\sigma^\mu \bar{\epsilon})_\alpha \partial_\mu \itindex{\varphi}{elim} + 2
  c \tr \lambda_\alpha (\epsilon \sigma^\mu D_\mu \bar{\lambda}) + 2c \tr
  \lambda_\alpha (\bar{\epsilon} \bar{\sigma}^\mu D_\mu \lambda)\ , \medsp
\label{eq:SYM5.1}
  \delta \itindex{F}{elim} &= - i \bar{\epsilon} \bar{\sigma}^\mu \partial_\mu
  \itindex{\psi}{elim}\ .
\end{align}
As was expected from the analysis of the previous section, supersymmetry does
not close on the operators in $S$ once the auxiliary field has been
eliminated. This raises the question about the meaning of the effective action
\eqref{eq:SYM3}. This ansatz can be justified in the special case
of SYM at least within a restricted range. Indeed it follows from the explicit
expression for the supercurrent in \eqref{eq:SYM1}
\begin{equation}
  \label{eq:SYM6}
  V_{\alpha \dot{\alpha}} = \inv{\cg} \tr W_\alpha e^{-V}
  \bar{W}_{\dot{\alpha}} e^V
\end{equation}
that the new contributions, which appear on the right hand side of eq.\
\eqref{eq:SYM5}, are components of $\bar{D}^{\dot{\alpha}} V_{\alpha \dot{\alpha}}$. Equation \eqref{eq:SYM1}
then suggests the operator relations
\begin{align}
  \label{eq:SYM7}
  \lambda_\alpha (D_\mu \lambda \sigma^\mu)_{\dot{\alpha}} &= \sigma^\mu_{\alpha
  \dot{\alpha}} \partial_\mu \itindex{\varphi}{elim}\ ,\medsp
  \label{eq:SYM8}
  \lambda_\alpha (D_\mu \bar{\lambda} \bar{\sigma}^\mu)^{\beta} &= i
  \delta^\beta_\alpha \itindex{F}{elim}\ .
\end{align}
Classically, which means $\delta_\alpha \rightarrow 0$ in \eqref{eq:SYM1}, the right hand side of the eqs.\ \eqref{eq:SYM7} and
\eqref{eq:SYM8} is zero and thus the Lagrangian multiplet transforms with the
linear transformations of a chiral field even after the
auxiliary field $D$ has been eliminated! At the quantum level the naive
application of eqs.\ \eqref{eq:SYM7} and
\eqref{eq:SYM8} would lead to the conclusion that
\begin{equation}
  \label{eq:SYM9}
  \delta (\itindex{\psi}{elim})_\alpha = \bigl(\epsilon_\alpha \itindex{F}{elim} -
  i (\sigma^\mu \bar{\epsilon})_\alpha \partial_\mu
  \itindex{\varphi}{elim}\bigr)\bigl( 1 + \inv{12} \frac{\beta}{g^3}\bigr)\ .
\end{equation}
However, one should be more careful with the interpretation of these
calculations. In contrast to the Wess-Zumino model a naive elimination of the
auxiliary field $D$ in SYM does not make sense at the quantum level. In
Wess-Zumino gauge supersymmetry transformations mix with the gauge symmetries
and thus the BRST construction is more involved
\cite{Howe:1990pz,White:1992ai,Maggiore:1996gr,hollik99}.
Nevertheless, the simple considerations of this section show that --due to the
basic structure of the anomalous current conservation-- the supersymmetry
transformations of the components of $S$ close on this set of operators even
if the auxiliary field $D$ has been eliminated. The true transformation
rules for the latter case demand a careful quantization in Wess-Zumino
gauge, which should follow along the lines of refs.\ \cite{rupp01,rupp01:2}. Though it is
expected that this leads to a modification of our result, the
appearance of quantum deformations of the supersymmetry transformations (as in \eqref{eq:SYM9}) cannot be
excluded. The exact answer and its influence on the anomaly structure
\eqref{eq:SYM1} must be left open in the present work.

We conclude this section with the two main statements, which follow from the
considerations made so far:
\begin{enumerate}
\item At least for vanishing sources $J$, supersymmetry appears to close on the
  components of the composite superfield $S$. This is obvious in the superspace
  representation, but by means of the anomalous current
  conservation it holds after the elimination of $D$ as well.
\item This seems to justify the ansatz for an effective
  action in terms of the field $\eff{S}$ alone. Nonetheless the
  calculation clearly shows that no component of $\eff{S}$ is an auxiliary
  field. Thus the effective potential in \eqref{eq:SYM3} must be bounded from
  below in \emph{all} fields and at the same time derivatives on $\eff{F}$ are
  necessary as well. This is impossible if $K(\eff{S},\eff{\bar{S}})$ is a
  K\"ahler manifold, but requires a more general ansatz
  \cite{Shore:1983kh,bergamin02:1,Bergamin:2003ub,portmann03}. 
\end{enumerate}

\section{Soft Supersymmetry Breaking in SYM}
There remains a surprising characteristic of the effective action found in the
previous section: This action is formulated in terms of two complex scalar
fields $\varphi$ and $F$ and one Majorana spinor $\psi$. As both scalars must
be interpreted as propagating degrees of freedom, on-shell this action has
four bosonic and two fermionic degrees of freedom. This seems to contradict supersymmetry, where the number of bosonic and fermionic
degrees of freedom should be equal. We are not able to present a conclusive
solution to this question, but the study of softly broken supersymmetry shows that it must be
seen in a larger context than discussed so far.

Softly broken supersymmetry is obtained in the formulation of the previous
section by taking a non-zero, constant limit of the source $J$. The complete classical
action from \eqref{eq:SYM2} with $J$ as defined below that equation takes the form 
\begin{equation}
  \label{eq:SYM10}
  \Ltext{SYM} = \intd{\diff{^2 \theta}} (\tau_0 + J) \tilde{S} + \hc =  (\tau_0+\tau) \tilde{F} - \eta \tilde{\psi} - 2 m
  \tilde{\varphi} + \hc\ .
\end{equation}
The contribution $\propto m$ is normalized in such a way that it leads to the
standard mass term for the gluino, $\tau_0$ is the bare coupling constant of
eq.\ \eqref{eq:SYM2}.

For $J = 0$ the global
symmetries are represented by the anomalous current conservation
\eqref{eq:SYM1}. In the presence of a non-vanishing source $J$ it is
advantageous to redefine the current $V_{\alpha \dot{\alpha}}$ of eq.\
\eqref{eq:SYM6} as
\begin{equation}
  \label{eq:SYM11}
    V_{\alpha \dot{\alpha}} = \inv{2 \cg} (\tau_0 + \bar{\tau}_0)\tr W_\alpha e^{-V}
  \bar{W}_{\dot{\alpha}} e^V\ .
\end{equation}
Then the breaking terms take the form \cite{bergamin01:2}
\begin{equation}
  \label{eq:SYM12}
  \bar{D}^{\dot{\alpha}} V_{\alpha \dot{\alpha}} = - 2 (D_\alpha J) \tilde{S}
  + \inv{\tau_0 + \bar{\tau}_0} (\bar{D}^{\dot{\alpha}} \bar{J}) V_{\alpha
  \dot{\alpha}} + \frac{J + \bar{J}}{\tau_0 + \bar{\tau}_0}
  \bar{D}^{\dot{\alpha}}
  V_{\alpha \dot{\alpha}} + \delta_\alpha\ .
\end{equation}

The physically most interesting situation is certainly soft supersymmetry
breaking due to a finite mass of the gluino, which means $J = - 2 \theta^2 m$,
whereby $m$ is constant. As $m$ does not couple to $\tilde{\psi}$ in
\eqref{eq:SYM10} the equation of motion for $D$ still reads $D = 0$. Thus the
components of $\itindex{S}{elim}$ still transform according to
\eqref{eq:SYM4}-\eqref{eq:SYM5.1}. But due to the right hand side of eq.\
\eqref{eq:SYM12}, the modification of the transformation law of $\tilde{\psi}$
in \eqref{eq:SYM5} do no longer vanish even setting $\delta_\alpha = 0$. Instead the
eqs.\ \eqref{eq:SYM7} and \eqref{eq:SYM8} are changed according to
\begin{align}
  \label{eq:SYM13}
  \lambda_\alpha (D_\mu \lambda \sigma^\mu)_{\dot{\alpha}} &= - i
  \frac{2m}{\tau_0 + \bar{\tau}_0} R_{\alpha \dot{\alpha}} + \mbox{quantum corrections}\ ,\medsp
  \label{eq:SYM14}
  \lambda_\alpha (D_\mu \bar{\lambda} \bar{\sigma}^\mu)^{\beta} &= - \half{i}
  \delta_\alpha^\beta m \itindex{\tilde{\varphi}}{elim} + \mbox{quantum corrections}\ ,
\end{align}
with the $R$ current $R_{\alpha \dot{\alpha}}$ defined as the lowest component
of \eqref{eq:SYM11}. This leads to the transformation rule
\begin{equation}
  \label{eq:SYM15}
  \delta (\itindex{\tilde{\psi}}{elim})_\alpha = \epsilon_\alpha
  \bigl(\itindex{\tilde{F}}{elim} + \half{1} m \itindex{\tilde{\varphi}}{elim} \bigr) -
  i (\sigma^\mu \bar{\epsilon})_\alpha \partial_\mu
  \itindex{\tilde{\varphi}}{elim} + \frac{m}{4(\tau_0 + \bar{\tau}_0)} R_{\alpha
  \dot{\alpha}} \bar{\epsilon}^{\dot{\alpha}} + \mbox{quantum corrections} \ .
\end{equation}
In eq.\ \eqref{eq:SYM15} the deformation of the transformation rule from superspace is
not just a result of the quantization, but contains classical contributions as well.

Of course any deformation of the superspace transformation rules immediately
raises the question how to construct an invariant action respecting the
``on-shell'' symmetry transformations. This certainly applies to the result of
eq.\ \eqref{eq:SYM15} as well, but, even more important, this equation shows
that an effective action in terms of $\eff{S}$ alone is not complete. There
exists no operator identity that would transform the $R$ current in eq.\
\eqref{eq:SYM15} into an expression in terms of the components of $\itindex{S}{elim}$ and
consequently supersymmetry does not close on the operators of
$\itindex{S}{elim}$, at least if the gluino mass does not vanish\footnote{One might
  wonder why we still insist on correct supersymmetry transformation rules,
  once this symmetry is broken anyway. However, it is important to realize
  that the action \eqref{eq:SYM10} \emph{is} invariant under supersymmetry, if
the sources $J$ are transformed as well. This conserved supersymmetry has its
own current and the result \eqref{eq:SYM15} could have been derived therefrom
as well. The shortcut of using the partially conserved current is allowed, as
$\itindex{\tilde{\psi}}{elim}$ does not depend on the sources and moreover we think
that the breaking terms are more transparent in this way.}.

This result demonstrates that an equal number of fermionic and bosonic degrees
of freedom can be found in a larger context, only. To include besides the
anomaly multiplet $S$ the supercurrent $V_{\alpha \dot{\alpha}}$ as
field relevant for the effective action, was proposed in \cite{Shore:1983kh}
already. But the ansatz in ref.\ \cite{Shore:1983kh} led to complicated differential equations in
superspace, which could not be solved. Within the perturbative framework, some
relevant results have been derived in \cite{kraus01,kraus01:3,Kraus:2002nu}. It is not the purpose of this paper to
investigate the technical details of such a formulation, but we conclude
this section with some comments on the difficulties in the construction of a
consistent effective action:
\begin{itemize}
\item $S$ and $V_{\alpha \dot{\alpha}}$ do not restore an equal number of
  bosonic and fermionic degrees of freedom.
\item For non-vanishing gluino mass supersymmetry closes on the operators in
  $S$ and $V_{\alpha \dot{\alpha}}$, but this is not true for non-vanishing
  fermionic source $\eta_\alpha$. We were not able to find a definite
  interpretation of the rather complicated formulas, which follow in that case.
\item Both points indicate that an even more general field and source content is
  necessary: $J^{\alpha \dot{\alpha}}$ needed to couple the supercurrent
  $V_{\alpha \dot{\alpha}}$ to the SYM action contains a source for the
  energy-momentum tensor. Thus the system must be embedded in a non-trivial
  supergravity background together with all currents of the
  super-conformal structure. Only within this maximal set of gauge invariant
  and supersymmetry covariant fields a resolution of the supersymmetry
  representation on composite effective fields can be expected. On the way
  there, the fate of the different constraints on the superfields, which are
  crucial for the formulation of the classical theory,
  must be considered at the quantum level. This amounts to study an anomaly
  structure within a more general context than done in \eqref{eq:SYM1}
  including eventual non-vanishing $\delta_\alpha$.
\end{itemize}

\section{Conclusions}
The elimination of auxiliary fields is an important procedure in all
supersymmetric theories that allow a superspace formulation. At the level of
the classical Lagrangian its meaning is obvious: Elimination of an auxiliary
fields means to impose a constraint, which holds the field at the extremum
(maximum) of its potential. The physics of the theory do not change under this
elimination.

At the level of effective actions the situation is more involved: There exist
no separate constraints for the effective fields, but an eventual elimination
must follow from the constraints on the fundamental fields. The result of
these considerations depends on the details of the effective fields:
\begin{itemize}
\item If the effective fields are simple operators, i.e.\ the sources couple
  to field monomials, the elimination of the fundamental auxiliary fields
  induces the functional restriction of the effective auxiliary fields. The
  effective action as well as the supersymmetry transformations may be written
  in terms of the remaining fields.
\item If the effective fields are composite operators, the elimination of the
  fundamental auxiliary field leads to a change in the field content
  of the effective fields as well. But none of the resulting
  effective fields may be written as a function of the other ones. Thus there
  exists no concept of the elimination of fields in an effective action solely
  written in terms of composite effective fields.

  The supersymmetry transformations lead to an even more stringent condition:
  The ``on-shell'' supersymmetry transformation do not close on a subset of
  the effective fields, which confirms the above statement. But in the generic
  case they do not even close on the composite fields stemming from those
  superfields, which were used to define the effective action. If
  this happens the formulation in superspace must be inconsistent
  as well, as it must be physically equivalent to the non-linear ``on-shell''
  formulation. 
\end{itemize}

In the second part of the paper we analyzed the effective action of $N=1$ SYM
from this point of view:
\begin{itemize}
\item At the classical level ($\delta_\alpha = 0$ in eq.\ \eqref{eq:SYM1}) the
  formulation of the effective action in terms of $S$ alone appears to be
  consistent within a restricted range:
  \begin{enumerate}
  \item For vanishing sources supersymmetry closes even after the elimination
    of $D$ on the three operators in $S$. The supersymmetry transformations do
    not experience any change by this elimination. Clearly, no field in $S$
    can be treated as an auxiliary field.
  \item Softly broken supersymmetry, which is equivalent to a non-zero but
    constant source, cannot be described in terms of $S$ alone. E.g.\ a soft
    gluino mass induces supersymmetry transformations into the $R$
    current. Thus at least the supercurrent must be coupled with a source as
    well. This dictates to consider the model in a supergravity background, as
    it contains a source for the energy-momentum tensor.
  \end{enumerate}
\item At the quantum level a deformation of the classical ``on-shell''
  supersymmetry transformations must be expected, even without soft
  supersymmetry breaking.
\end{itemize}
These important characteristics are certainly a serious challenge for our
present knowledge about $N=1$ SYM theory. To study its consequences
necessitates investigations in many directions, esp.\ the results will never be found
within the purely holomorphic sector of the effective action. 

\subsection*{Acknowledgement}
The authors would like to thank E. Kraus and Ch.\ Rupp for important
discussions. One of us (L.B.)\ would like to thank J.-P. Derendinger,
U. Ellwanger and E.\ Scheidegger for interesting comments. Important parts of
this work have been developed on the occasion of a workshop at the
International Erwin Schr\"{o}dinger Institute. This work has been
supported by the Swiss National Science Foundation (SNF) and
the Austrian Science Foundation (FWF) project P-16030-N08.

\providecommand{\href}[2]{#2}\begingroup\raggedright\endgroup
\end{document}